\documentclass[aps, prl,reprint, groupedaddress, showpacs, showkeys,twocolumn]{revtex4-1}
\usepackage{graphicx}
\usepackage{amsmath, color, ulem}
\newcommand{\dpartial}[2]{\frac{\partial #1}{\partial #2}}
\newcommand{\ve}[1]{{\bf #1}}

\usepackage{soulutf8}
\usepackage{upgreek}
\usepackage{ulem}

\begin{document}


\title{Magnetic field control of the spin Seebeck effect}

\author{Ulrike Ritzmann}
\author{Denise Hinzke}
\author{Andreas Kehlberger}
\author{Er-Jia Guo}
\author{Mathias Kl\"aui}
\author{Ulrich Nowak}
\affiliation{Department of Physics, University of Konstanz, D-78457 Konstanz, Germany}
\affiliation{Institute of Physics, Johannes-Gutenberg University Mainz, 55099 Mainz, Germany}

\date{\today}

\begin{abstract}
The origin of the suppression of the longitudinal spin Seebeck effect by applied magnetic fields is studied. We perform numerical simulations of the stochastic Landau-Lifshitz-Gilbert equation of motion for an atomistic spin model and calculate the magnon accumulation in linear temperature gradients for different strengths of applied magnetic fields and different length scales of the temperature gradient. We observe a decrease of the magnon accumulation with increasing magnetic field and we reveal that the origin of this effect is a field dependent change of the frequency distribution of the propagating magnons. With increasing field the magnonic spin currents are reduced due to a suppression of parts of the frequency spectrum. By comparison with measurements of the magnetic field dependent longitudinal spin Seebeck effect in YIG thin films with various thicknesses, we find that our model describes the experimental data very well, demonstrating the importance of this effect for experimental systems. 
\end{abstract}

\pacs{75.30.Ds, 75.40.Mg, 75.76.+j} 

\maketitle

Spin caloritronics is an emerging research field promising spintronic devices with new functionalites, which rest on the combined transport of heat and spin \cite{Bauer_2012, Boona_2014b}. A key tool that stimulated this field is the inverse spin Hall effect \cite{Saitoh_2006, Valenzuela_2006}, which allows one to detect pure spin currents due to a transformation into measurable charge currents via spin orbit coupling. Using this indirect measurement technique Uchida et al. measured the so-called spin Seebeck effect (SSE) \cite{Uchida_2010a}. The measurements showed that in the ferromagnetic insulator YIG a pure spin current is created due to an applied temperature gradient.

These findings triggered a variety of further studies in different groups \cite{Weiler_2012, Kehlberger_2013, Schreier_2013, Kikkawa_2013, Kikkawa_2013b, Schreier_2013b, Agrawal_2014,Roschewsky_2014, Vlietstra_2014, Uchida_2014} to investigate the origin of the detected spin currents. Interface effects due to a proximity effect which creates a magnetization in the normal metal were discussed \cite{Huang_2012}, but various groups showed that the effect in the YIG/Pt system cannot be responsible for the measured signals \cite{Kehlberger_2013, Kikkawa_2013, Kikkawa_2013b, Vlietstra_2014}. A first theoretical description of the magnonic spin Seebeck effect was developed by Xiao et al. \cite{Xiao_2010}. With a two temperature model including the local magnon and phonon temperatures the measured spin Seebeck voltage was shown to originate from the local difference between magnon and phonon temperature at the Pt interface caused by thermal spin pumping. Later it was shown by spin model simulations that such a 
temperature difference can be caused by magnons traveling from the hotter towards the colder part of the sample \cite{Ritzmann_2014}. By studying the thickness dependence of the SSE for thin YIG films, Kehlberger et al.~\cite{Kehlberger_2013} found an increasing spin Seebeck signal saturating above a critical length scale. This length scale can be referred to as the mean magnon propagation length. Similarly Agrawal et al. measured the time evolution of the SSE in the sub-microsecond regime and explained their time dependent increase of the SSE by a dependence on the characteristic length scale of magnon propagation of a few hundred nm \cite{Agrawal_2014}.

Further theoretical studies focused on the magnonic origin of the SSE \cite{Hoffman_2013, Etesami_2014} also predicting thickness dependent effects. Hoffman et al. \cite{Hoffman_2013} studied the SSE analytically within the framework of the stochastic Landau-Lifshitz-Gilbert equation. They assume that mainly magnons with $\hbar\omega\approx k_{\rm B}T$ contribute to the effect, but Boona et al.~proposed that also sub-thermal magnons with lower energy can contribute or even dominate \cite{Boona_2014}. To reveal, which magnons contribute to the effect, field - dependent measurements that tailor the magnon spectrum are a key. First such measurements were made available in  a recent publication by Kikkawa et al.~\cite{Kikkawa_2013} who studied the reduction of the SSE in the regime of high magnetic fields and report a suppression of the measured SSE signal. However no theoretical explanation was given that reproduces the results completely and allows for a direct comparison of experiment and theory.

In this paper, we study the influence of external magnetic fields on the SSE. Applying external fields manipulates the frequency distribution of the propagating magnons by increasing the frequency gap $\omega_{\rm min}$ of the frequency spectrum. Therefore, the number of propagating magnons can be reduced by increasing the external magnetic field. We present numerical simulations of the magnon accumulation in linear temperature gradients and its dependence on external magnetic fields. For this study an atomistic spin model is used, which provides realistic spin wave spectra beyond parabolic approximation and without any artificial cut-off due to discretization effects. To understand the underlying physics, an analytical model is derived that explains the effect of the magnetic field within the framework of linear spin wave theory. Finally, our theoretical work is directly compared with field dependent SSE measurements of YIG thin films with different thicknesses, which cover a large range across the critical 
length scale showing good agreement.
 
For the numerical simulations we use an atomistic spin system consisting of localized spins with normalized magnetic moment $\ve{S}_i={\boldsymbol \mu}_i/\mu_{\rm s}$ on a cubic lattice. Our model Hamiltonian $\mathcal{H}$ includes  Heisenberg exchange interaction for nearest neighbors with exchange constant $J$, a uniaxial anisotropy with an easy-axis in $z$-direction with anisotropy constant $d_z$, and an external field $\ve{B}=B_z\ve{e_z}$ parallel to the easy axis,
\begin{align} 
    \label{hamiltonian}
    \mathcal{H}=-\frac{J}{2} \sum_{<i,j>}{\ve{S}_i\ve{S}_j}-d_z\sum_i{\big(S_{i}^z\big)^2}-\mu_{\rm s}B_z\sum_i{S_i^z}\;\mbox{.}
  \end{align} 
The dynamics of each single spin are described by the stochastic Landau-Lifshitz-Gilbert (LLG) equation,
\begin{align}
  \label{LLG}
  \dpartial{\ve{S}_i}{t}=-\frac{\gamma}{\mu_{\rm s}(1+\alpha^2)} \ve{S}_i\times\left(\ve{H}_i+\alpha\left(\ve{S}_i\times\ve{H}_i\right)\right)\mbox{,}
\end{align}
consisting of a precession around its effective field $\ve{H}_i$ and a phenomenological damping with damping constant $\alpha$ \cite{Landau_1935, Gilbert_1955}. $\gamma$ is the gyromagnetic ratio and the effective field $\ve{H}_i$ is given by
 \begin{align}
    \label{Heff}
    \ve{H}_i=-\dpartial{\mathcal{H}}{\ve{S}_i}+\boldsymbol{\zeta}_i(t)\;\mbox{.}
  \end{align}
The temperature is included as additional noise term $\boldsymbol{\zeta}_i(t)$ in the effective field $\ve{H}_i$ fulfilling 
\begin{align}
 \left\langle\boldsymbol{\zeta}(t)\right\rangle=0 \qquad \left\langle\zeta_i^{\eta}(0)\zeta_j^{\theta}(t) \right\rangle=\frac{2k_{\rm B}T_{\rm p}\alpha\mu_{\rm s}}{\gamma}\delta_{ij}\delta_{\eta\theta}\delta (t)\mbox{.}
\end{align}
The dynamics of the systems are studied by numerical integration of these equations using the Heun-method \cite{Nowak_2007}.

\begin{figure}[tb]
 \includegraphics[width=0.49\textwidth]{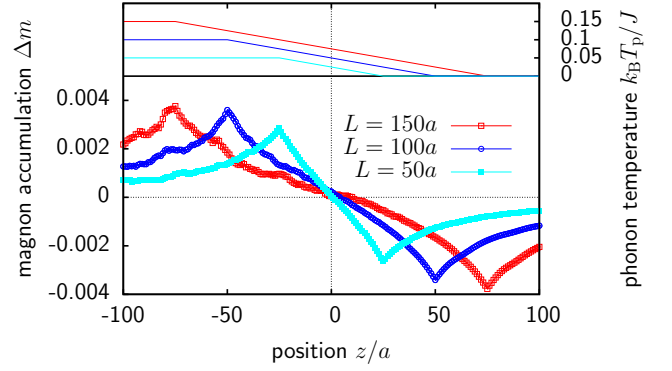} 
 \caption{\label{akkumulation} Local magnon accumulation $\Delta m$ profile with the space position $z$ in units of the lattice constant $a$ for different thicknesses $L$.}
\end{figure}

In the systems considered, the temperature is spatially dependent, including a linear temperature gradient in $z$-direction with a constant slope over a distance $L$ and two heat baths at the two ends of the system as shown in FIG. \ref{akkumulation}. The dimension of both heat baths is chosen to be large enough to minimize finite size effects  in the area of the gradient. The given temperature profile describes the temperature of the phononic heat bath and it is assumed that the phonon temperature remains constant during the simulation.

We study the manipulation of the magnon accumulation by applying external magnetic fields. For this purpose, we simulate a system with $8\times8\times512$ spins including an easy-axis parallel to $z$-direction with $d_z=10^{-3}$, a damping constant of $\alpha=0.01$ and apply an external field in the $z$-direction. Linear temperature gradients with variable lengths $L$ excite net magnonic spin currents due to a non-equilibrium of the local magnonic density of states. In the hotter area more magnons exist than in the colder area and therefore more magnons propagate towards the colder region than the other way round, leading to a net magnon current from hot to cold.
Since we aim to describe spin accumulation in samples with thin film geometry we will call the spatial extension of the gradient, $L$, from now on the thickness.

After an initial relaxation the system reaches a quasi static state where we calculate the local magnetization profile $m(z)$ as an average over time and over the $x$-$y$-planes. The magnon accumulation can be defined as the difference of the local magnetization to its equilibrium value related to the local phonon temperature,
\begin{align} 
 \Delta m(z)=m(z)-m_{\rm eq}(T_{\rm p}(z))\mbox{.}
\end{align}%

FIG. \ref{akkumulation} shows exemplary the resulting magnon accumulation for an external field $B_z=0.1J/\mu_{\rm s}$ and a temperature gradient of $\nabla_zT=10^{-3}J/(k_{\rm B}a)$ for various thicknesses $L$. As shown in similar simulations by Kehlberger et al. the magnon accumulation has two extrema at the ends  of the temperature gradient and their values increase with increasing thickness up to a characteristic length scale above which the magnon accumulation saturates. This characteristic length scale can be referred to as the mean magnon propagation length $\xi_{\rm avg}$ \cite{Kehlberger_2013}.

\begin{figure}[tb]
 \includegraphics[width=0.49\textwidth]{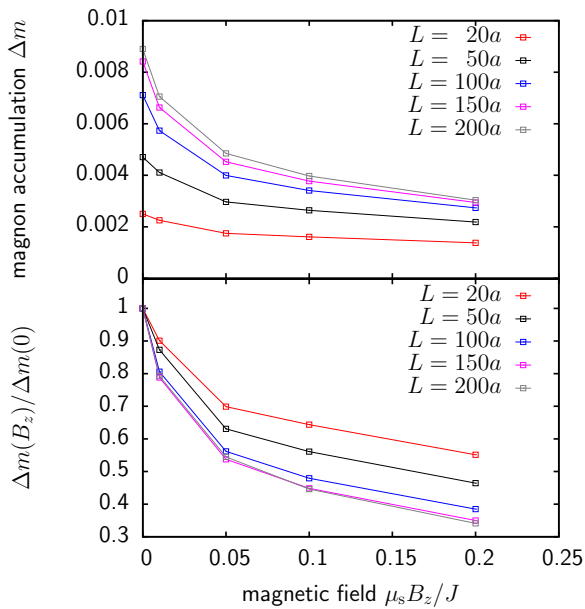} 
 \caption{\label{saettigung} Upper part: Magnon accumulation at the cold end of the temperature gradient versus applied magnetic field $B_z$ for different thicknesses with $\nabla T=10^{-3}k_{\rm B}T/(Ja)$, $\alpha=0.01$, $d_z=10^{-3}J$. Lower part: Corresponding normalized magnon accumulation $\Delta m(B)/\Delta m(0)$.}
\end{figure}

In the upper part of FIG. \ref{saettigung}, the extreme value of the magnon accumulation at the cold end of the temperature gradient is shown versus applied magnetic field $B_z$ for different thicknesses. The magnetic fields used in these simulations corresponds to magnetic fields of the order of 80\,T leading to strong effects. In all cases one can see that the magnon accumulation decreases with increasing field, but the observed suppression of the magnon accumulation is thickness dependent. For thicker films, larger than the mean magnon propagation length, the field suppression of the magnon accumulation is stronger than for thinner films where  the suppression effect shows only a weak thickness dependence. 
This can be seen in the lower part of FIG. \ref{saettigung} where the normalized magnon accumulation $\Delta m(B_z)/\Delta m(0)$ is shown for various thicknesses. 

The magnon accumulations at both, the hot and the cold end of the temperature gradient, are nearly proportional to the temperature difference between the magnonic and phononic subsystems \cite{Ritzmann_2014}. It was shown by Xiao et al. that this temperature difference scales linearly with the spin Seebeck voltage, which is measured in  experiments \cite{Xiao_2010}. Hence, one can expect that the magnonic contribution to the SSE can be suppressed by magnetic fields. Interestingly, this behavior shows a thickness dependence with the influence of the field decreasing for thinner films, an effect that we will discuss later on.

\begin{figure}[tb]
 \includegraphics[width=0.49\textwidth]{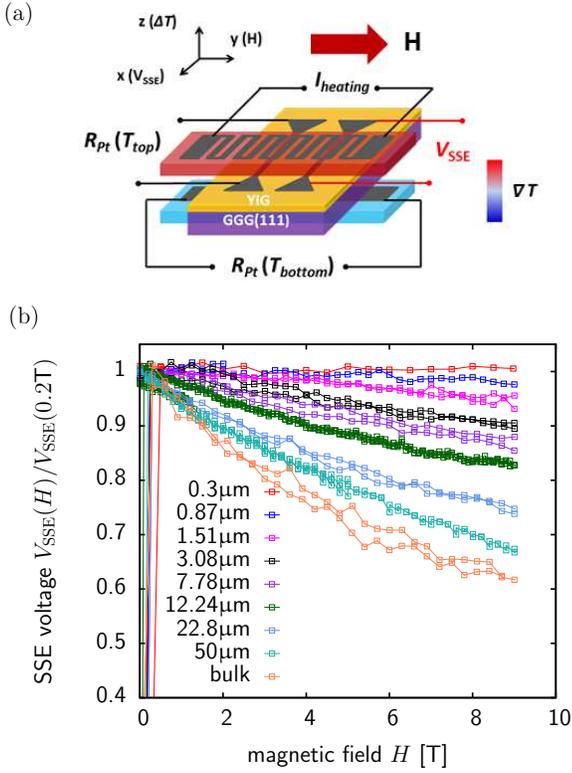} 
 \caption{\label{experiment}
 (a) Schematic diagram of sample structure and measurement setup. (b) Normalized spin Seebeck voltage $V_{\rm SSE}(H)/V_{\rm SSE}(0.2\,T)$ of Pt/YIG structures composed of YIG single crystal and selected YIG thin films with thickness ranging from 0.3 to 50$\,\upmu$m.}
\end{figure}

First, in order to verify our theoretical model, the high magnetic field dependence of the SSE is measured using Pt/YIG hybrids. The samples used in the present study are (111)-oriented YIG slabs (5$\times$10$\times$1 mm$^3$) and YIG thin films grown on Gd$_3$Ga$_5$O$_{12}$ (GGG) substrate (5$\times$10$\times$0.5\,mm$^3$) by liquid phase epitaxy. The thickness of the YIG thin films is varied from 0.3\,$\upmu$m to 50\,$\upmu$m. 5.5 nm-thick Pt layers are deposited on the YIG surfaces by dc-magnetron sputtering and further patterned into strips (length of 4 mm and width of 100 $\upmu$m) by optical lithography and argon ion beam milling. FIG.~\ref{experiment}(a) shows a schematic diagram of the prepared sample and measurement setup. We adopt the longitudinal configuration to determine the SSE. The sample is sandwiched between the resistive heater and thermal sensor, further mounted onto a copper heat sink then put into the cryostat. Each functional layer is structured on a Al$_2$O$_3$ substrate, which prevent 
an electrical short circuit the individual elements another, while ensuring maximal vertical temperature transport. Furthermore a thermal grease is used for the mounting of the elements, which ensures good thermal connection at the interfaces. A temperature gradient, $\nabla T$, can be generated by the attached resistive heater and be varied by simply changing the heating currents. A spin current is thermally generated in the Pt layer along the direction of thermal gradient ($z$-axis), and further converted into an electric field due to the inverse spin Hall effect. The SSE can be detected electrically by measuring the voltage drop $V_{\rm SSE}$ at the two ends of the Pt strip (along $x$-axis). The benefit of our measurement setup is that the Pt strips on the top and bottom surfaces of the samples can be utilized as an excellent resistance-temperature detector to determine the temperature differences across the hybrids precisely. In our room-temperature experiments we keep the cryostat at a fixed temperature 
of 300\,K and a heating current in our resistive heater of $I_{\rm heat} = 9$\,mA results in the temperature difference $\Delta T$ of roughly $4.2 \pm 0.2$\,K for our measurements. An external magnetic field $H$ is applied perpendicular to the Pt strip along $y$-axis. The maximum magnetic field is up to 9\,T, which is orders of magnitude higher than the coercivity of YIG, ensuring that the magnetization of the YIG slab (or YIG thin films) is well aligned along the field direction. To exclude proximity effects from the Pt/YIG interface as the origin of the observed field suppression of the SSE,  the magnetoresistance of the Pt layer is monitored, which shows no noticeable change beyond the saturation field of magnetization of the YIG film.

The $V_{\rm SSE}$ is recorded as a function of magnetic fields. The SSE signal appears when a magnetic field is applied and flips its sign when the field reverses. We find the magnitude of the SSE signal suppressed by high magnetic fields at room temperature. The maximum value of $V_{\rm SSE}$ is found at the smallest field interval of 0.2\,T. The thickness dependent SSE coefficient $\sigma_{\rm SSE}=V_{\rm SSE}(0.2\,T)/\Delta T$ exhibits a similar trend to the one observed in our previous work: the $\sigma_{\rm SSE}$ is enhanced with increasing film thickness and saturates above a characteristic length, demonstrating the bulk origin of magnonic spin current \cite{Kehlberger_2013}. FIG.~\ref{experiment}(b) shows examples of the normalized spin Seebeck voltages $V_{\rm SSE}(H)/V_{\rm SSE}(0.2\,T)$ for selected Pt/YIG hybrids. Obviously, the suppression effect is reduced dramatically with decreasing film thickness. The experimental results are in agreement with the numerical results using our proposed 
theoretical 
model showing a field suppression up to 40\% for the 1mm thick YIG slab under a magnetic field of 9\,T, while this value drops to only 0.1\% for 0.3$\,\upmu$m thick YIG films.

\begin{figure}[tb]
  \includegraphics[width=0.49\textwidth]{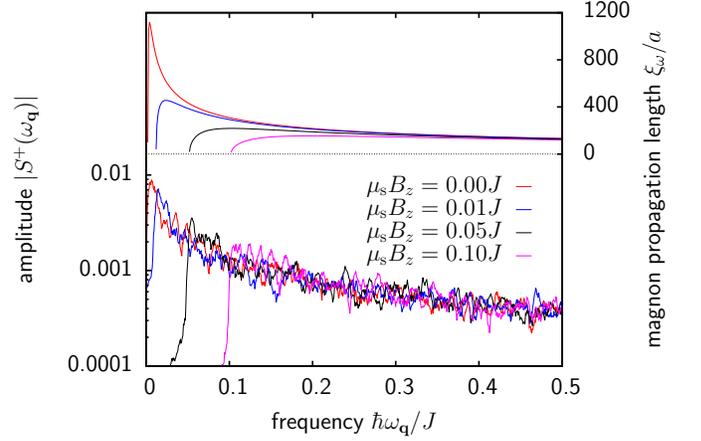}
 \caption{\label{reichweite} Magnon frequency distribution of propagating magnons at the cold end of the temperature gradient and the frequency dependent magnon propagation length for different magnetic fields.}
\end{figure}

To understand this effect at a fundamental level, we analyze it next based on an analytical linear spin wave theory. In particular we determine in this model the origin of the field suppression of the magnon accumulation as well as the thickness dependence of this effect. The magnonic dispersion relation of the simulated system can be calculated by solving the linearized LLG equation,
\begin{align}
 \hbar\omega_{\ve{q}}=2d_z+\mu_{\rm s}B_z+2J\sum_{\theta}(1-\cos(q_{\theta}a_{\theta})) \;\mbox{,}
\end{align}
where $\theta$ denotes the spatial coordinates $x,y,z$. The dispersion relation consists of a magnon frequency gap $\hbar\omega_{\rm min}=(2d_z+\mu_{\rm s}B_z)$ and a second term depending on the wave vector $\ve{q}$. The  frequency gap increases with increasing magnetic fields and, hence, parts of the spectrum are frozen out for higher magnetic fields. This effect can be seen in Fig.~\ref{reichweite}. By calculating a Fourier transformation of the magnon accumulation $\Delta m$ in the time domain, as used by Ritzmann et al. \cite{Ritzmann_2014}, at the cold end of the temperature gradient, we can calculate the frequency distribution of the magnons from our simulations. The shown frequency distributions for temperature gradients with width $L=200a$ for different magnetic fields shows a clear shift of the frequency gap and therefore a suppression of parts of the frequency spectra. This effect can explain the suppression of the accumulation due to applied magnetic fields.  

The frequency dependent magnon propagation length $\xi(\omega)$ is also shown in Fig.~\ref{reichweite}. It can be estimated using the lifetime $\tau=(\alpha\omega)^{-1}$ and the group velocity of the magnons \cite{Ritzmann_2014} leading to the expression: 
\begin{align}
  \frac{\xi_{\omega_{\ve q}}}{a}=\frac{\sqrt{J^2- \Big(\frac{1}{2}( \hbar \omega_q-\hbar\omega_{\rm min})-J \Big)^2}}{\alpha\hbar\omega_{\ve q}}
\end{align}
The maximum propagation length is given by
\begin{align}
 \frac{\xi_{\rm max}}{a}=\frac{1}{\sqrt{2}\alpha}\sqrt{\frac{J}{\hbar\omega_{\rm min}}}
\end{align}
and can be much larger than the mean propagation length of the magnons.

The thickness dependence of the field suppression can be explained by the frequency dependent magnon propagation length. Magnons with large propagation length will always reach the cold end of the temperature gradient. When the thickness is reduced, the contribution of those magnons is reduced. Since only these magnons are suppressed by the frequency gap, the field suppression is also reduced for low thicknesses. Note, that the equations above show that the same suppression effect would appear by modifying the anisotropy in the system, since the effect depends not only on the size of the magnetic field, but on the value of the frequency gap.

Furthermore, not only parts of the spectrum are suppressed, but also the frequency dependent magnon propagation length $\xi(\omega)$ is modified by applying higher magnetic fields. When the propagation length of the magnons is reduced, the number of magnons propagating through the system is smaller, leading also to a reduction of the magnon accumulation, showing that multiple effects contribute to this generic property of thermal magnon propagation in applied magnetic fields. 

In conclusion we have shown numerical simulations and experiments on the magnetic field suppression of the longitudinal SSE. Applying large magnetic fields can suppress the excited magnon spin currents by suppressing parts of the frequency spectrum. We find qualitative agreement between the performed simulation and experiments of the field dependence of the longitudinal SSE in YIG at room temperature for various thicknesses. The suppression is strongly dependent on the thickness of the sample and is practically vanishing for thin films since low frequency magnons play a minor role in thin films as compared to bulk materials. Furthermore, not only parts of the spectra are suppressed but the magnon propagation length given by the Gilbert damping is modified leading to a lower propagation length in systems with higher magnetic fields. Our analysis opens a new avenue to determining which magnons contribute to the spin current as a careful study of the field dependence would allow then to measure the intensities 
of the involved frequency as proposed by Boona et al. \cite{Boona_2014}. Furthermore, the field suppression open new opportunities for the control of the magnon flow and resulting SSE with external fields. During the preparation of this manuscript we became aware of two related experimental works in arxiv \cite{Kikkawa_2015, Jin_2015} supporting our findings.

\begin{acknowledgments}
 The authors would like to thank the {\it Deutsche Forschungsgemeinschaft} for financial support via SPP 1538 ``Spin Caloric Transport'' and the SFB 767 ``Controlled Nanosystem: Interaction and Interfacing to the Macroscale'' as well as the EU (IFOX, NMP3-LA-2012246102, INSPIN, FP7-ICT-2013-X 612759, MASPIC, ERC-2007-StG 208162).
\end{acknowledgments}

\bibliography{Quellen.bib}

\end{document}